\documentclass[final,5p,times]{elsarticle}

\usepackage{amsmath,amssymb,amsfonts}
\usepackage{graphicx}
\usepackage{bm}
\usepackage{physics}
\usepackage{ragged2e}
\usepackage{xcolor}
\usepackage{comment}
\usepackage{booktabs}
\usepackage{array}
\usepackage[table]{xcolor}
\usepackage{makecell}
\usepackage{subcaption}
\usepackage{xurl}

\usepackage{hyperref}
\hypersetup{
    colorlinks=true,
    linkcolor=blue,
    citecolor=blue,
    urlcolor=blue
}

\definecolor{lightgray}{gray}{0.9}

\begin{document}

\begin{frontmatter}

\title{Visibility graph-based characterization of extreme values in time series}
\author[inst1]{Juliane T. Moraes}
\author[inst2]{Lucas Lacasa} 
\author[inst1]{Cristina Masoller} 

\affiliation[inst1]{organization={Universitat Politècnica de Catalunya, Departament de Fisica, Rambla St. Nebridi 22},
            city={Terrassa 08222},
            country={Spain}}

\affiliation[inst2]{organization={Institute for Cross-Disciplinary Physics and Complex Systems (CSIC-UIB), Campus UIB, 07122},
            city={Palma de Mallorca},
            country={Spain}}

\begin{abstract}
{Complex dynamical systems often display extreme fluctuations of an observed variable that constitute significant deviations from the long-term average, and which are often associated with severe impacts on the system. By definition, extreme events are therefore usually explored from time series recordings. 
In this work, we 
characterize extreme values in time series using visibility graphs --a method that non-parametrically maps a time series of $N$ points onto a network of $N$ nodes, whose topological structure is known to inherit important characteristics of the original time series dynamics. Unlike threshold-based approaches, extreme values in this framework can be identified without the need 
of introducing external parameters and can be applied to time series generated by both, stationary and nonstationary processes. 
For stationary processes, we exploit a known property of visibility graphs in which the degree of a node is monotonically and nonlinearly related to the corresponding data value. This nonlinear amplification enhances the contribution of large values while suppressing noise, while the monotonic relationship enables a direct ranking of data points according to node degree. This procedure identifies not only 
global 
extreme values, but also, locally prominent 
ones. 
For nonstationary processes, 
the degree ranking in the visibility graph still provides a robust indicator of relative importance. We validate our findings with synthetic time series 
and with real climatological data. 
Our results show that extreme-value 
characterization in stationary time series is enhanced when combining standard methods with visibility-graph-based detection, whereas for nonstationary data --where 
conventional approaches are often ill-posed-- visibility graphs provide an effective alternative.
Additionally, we discuss how sub-sampling the time series using only peak values preserves the ability to identify extreme values 
while significantly reducing computational cost.
}

\end{abstract}


\end{frontmatter}

\section{Introduction}

{Extreme events (EEs) are ubiquitous in natural and artificial systems, and often constitute precursors to dynamic transitions toward different and potentially dangerous regimes \cite{book_kantz}. In our climate, examples of extreme events include extreme droughts that can precede desertification; in neural systems, large neuronal avalanches can trigger highly synchronized spike activity \cite{beggs2003neuronal}; in financial systems, a large fluctuation in the value of some stocks can generate enormous demand to buy or sell those stocks \cite{sornette2003critical}.}

{Extreme events are usually associated with extreme values (EVs) in the temporal evolution of a monitored variable, and EVs are identified by using two approaches: the threshold-based approach, in which each value of the time series that exceeds a certain threshold is considered an extreme value, and the interval-based approach, in which a time interval is defined and the extreme value is the highest value of the monitored variable in that time interval. The former approach implicitly assumes that the marginal distribution of the time series values is stationary and reasonably homogeneous, so that distribution outliers can be well-defined by comparing the value of a potential extreme with respect to the mean and variance of the underlying marginal distribution.}

{Examples of extreme events defined using the threshold-based approach are ocean rogue waves and optical rogue waves, which are extremely high waves \cite{book_rws, solli_2007}. The oceanographic community commonly uses as threshold the abnormality index, which is the ratio of wave height to the mean height of the highest one-third of the waves. In this way, any wave with an abnormality index $>2$ is considered a rogue wave (RW). An alternative definition of the threshold is in terms of the mean value and the standard deviation of the distribution of waves heights: any wave whose height exceeds the threshold, equal to the mean height plus a certain number of standard deviations, is considered a rogue wave \cite{rw_prl_2011}. Percentiles are also used to define the threshold that an extreme event should exceed. For example, extreme rainfall events have been defined as days with rainfall sums above the 95th percentile of wet days \cite{kurths_nature}. On the other hand, examples of extreme events defined in a time-interval are temperature extremes, such as the highest temperature in a particular month, year, decade, etc.}

{We note 
that traditional approaches for identifying EVs in time series do not always allow to identify high data values that are locally high, but not high enough to be cataloged to be EVs in terms of distribution outliers, nor the highest values in the pre-defined time interval analyzed. In other words, data points that are locally extreme --i.e. significantly higher that the neighboring data points-- may not be identified because they are not high enough (they are below the threshold) or because they are not the highest values in the time interval analyzed. However, in occasions these local EVs carry significant information. Indeed, 
local EVs can be 
as relevant as global EVs because they 
can trigger instabilities and regime transitions. 
{The timing of local and global EVs can be extremely relevant \cite{corral}.}
For example, a temperature of 38ºC in Spain might not constitute an outlier throughout the whole year, but if this temperature is reached in March, this is indicative of a 
dangerous climatological event. Likewise, a few days of rain can have a significant impact on an ecosystem after a long period of drought, even if on a global scale does not constitute by itself an extreme rainfall event.}

A second observation is that traditional characterization of extreme events via global EV detection implicitly assumes that the process is stationary --we need to be able to accurately compute the mean and standard deviation of the time series in order to evaluate when a data point is a distribution outlier--. However, many time series are notably nonstationary, such as temperature records in the face of climate change, or financial data. It is thus far from clear how to meaningfully define EVs in nonstationary time series, if at all possible.

{Here we address the two issues above by building on the visibility graph technique (VG) \cite{Lacasa2008, luque2009horizontal}, a well-known graph-based method for nonlinear time series analysis. The theory of visibility graphs has been fruitfully applied 
to characterize a variety of time series and their underlying processes in terms of topological properties (degree sequence, degree distribution, clustering, graph entropies, graph eigenvalues, graph motifs, etc) \cite{lacasa2009visibility, lacasa2010description, lacasa2012time, lacasa2018visibility, lacasa2015time, iacovacci2016sequential, luque2017canonical}, finding applications across disciplines including material science \cite{bohmer2024time}, fluid dynamics \cite{iacobello2021large}, laser dynamics \cite{prl_cris}, neuroscience \cite{sannino2017visibility}, finance \cite{zhuang2014time}, spreading processes \cite{moraes2023b,Moraes2025} and climate science \cite{zhang2024spatio, zhao2023visibility, beltramone2025detecting}. The VG technique is nowadays used as a feature extraction protocol for deep learning models thanks to their ability to capture relevant information of the time series structure 
\cite{zhang2022automated, xuan2022avgnet, aslan2024visgin, belhadi2025enhanced}. In a nutshell, VG 
combines network science \cite{Boccaletti_2006,Boccaletti_2016} and time series analysis \cite{kantz2003nonlinear}, 
in a powerful} graph-based time series analysis methodology \cite{review_kurths}.

Recent works have shown that VG can be used for extreme event detection, 
\cite{zhao2020extreme} via horizontal visibility graphs, \cite{zhang2023extreme} via VGs, or \cite{katsouda2025extreme} via so-called bar visibility graphs. In this work, we 
show that VGs provide a natural and non-parametric way to characterize not only global EVs but also local EVs which can be applied to both, stationary and nonstationary time series. 
Based on these insights, we develop a computationally efficient subsampling method capable of detecting local and global EVs, complementing traditional approaches.
We test our methodology on both synthetic and 
real time series and showcase its applicability across stationary and nonstationary processes.

The rest of this paper is organized as follows. In section \ref{sec:data} we describe 
the data analyzed, 
and in section \ref{sec:methods} we 
present the methodology used, alongside the theoretical foundation 
that supports and justifies how and why VGs properties are related to time series' local and global EVs. In section \ref{sec:results} we validate our methodology on four datasets, which include optical rogue waves, climatological data and benchmark synthetic data. Finally, in section \ref{sec:conclusion} we present the discussion and summarize our conclusions. 



\section{Data}\label{sec:data}

\begin{figure}[tb] 
\centering
\includegraphics[width=\linewidth]{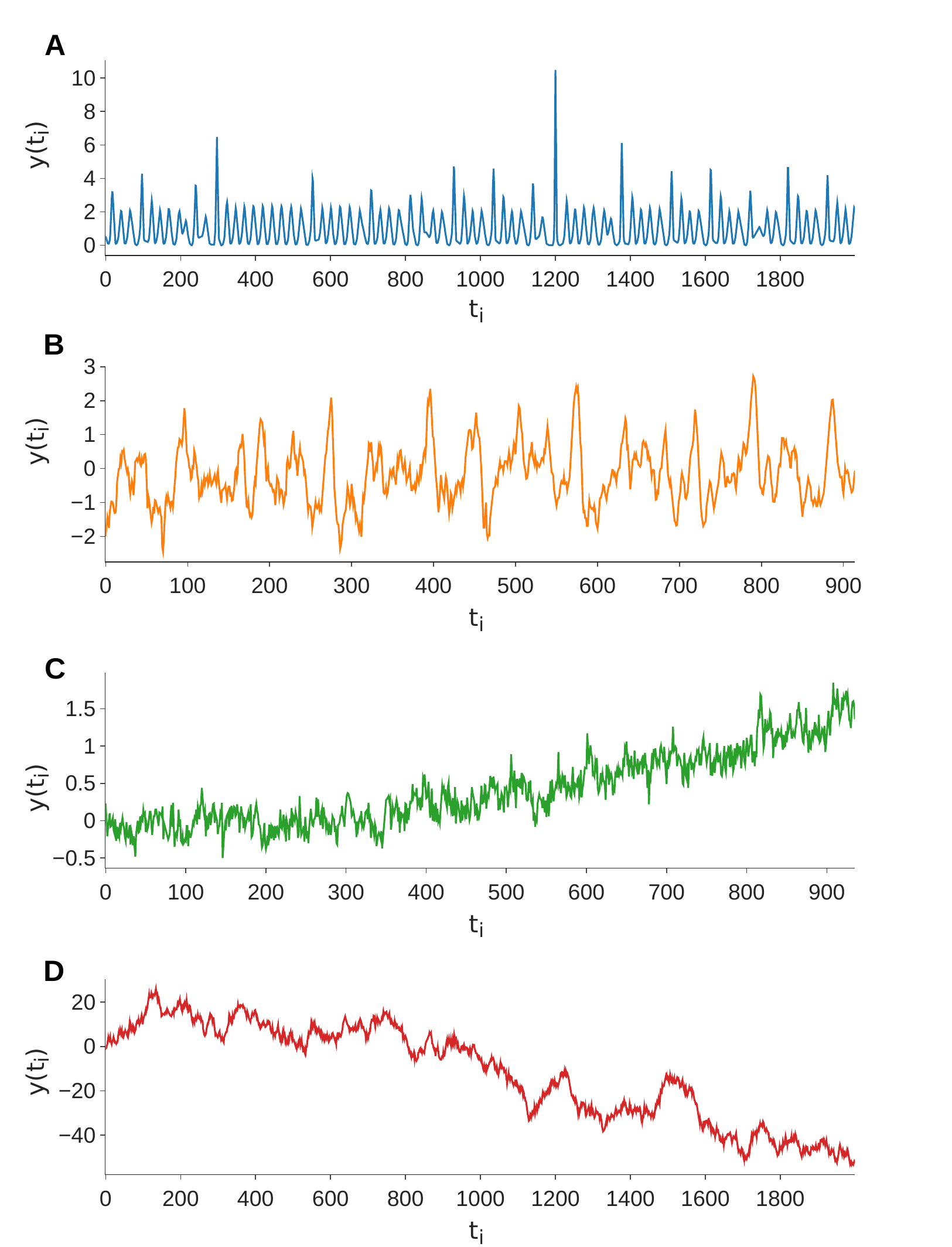} 
\caption{\justifying Time series analyzed in this work. (a) Simulated laser intensity, using model and parameters from \cite{rw_prl_2011}. (b) Niño 3.4 Sea SST Anomaly Index download from \cite{NOAA_Nino34_web}. (c) Global Mean Land/Ocean Temperature Anomaly Index download from \cite{NOAA_gmsst_web}. (d) Unbiased random walk displacement computed using $x_{t+1} = x_t+\xi$, with $\xi$ being uncorrelated Gaussian noise.  }
\label{fig:datasets}
\end{figure}

{In this work we analyze four time series that are generated by processes with different characteristics, stationary or not, some showing clear global extreme values and some showing the presence of local extreme values and the absence of a well-defined global extrema.} {The first time series,} 
{Fig. \ref{fig:datasets}(a),} {represents the intensity emitted by an optically injected semiconductor laser, and it was generated by simulating a simple deterministic rate-equation model \cite{rw_prl_2011}. The parameters used are those in Fig. 4 in \cite{rw_prl_2011}, and correspond to the optical rogue wave (ORW) regime, where occasionally the laser emits high pulses.} We use this time series as a paradigmatic case of a stationary time series generated by a deterministic, chaotic process with well-defined extreme values.

{The second time series analyzed is a well-known one that characterizes a global climate phenomenon: El Niño–Southern Oscillation (ENSO). We analyze the Niño 3.4 Sea Surface Temperature (SST) Anomaly Index, provided by the Physical Sciences Laboratory of the National Oceanic and Atmospheric Administration (NOAA). This index is derived from area-averaged SST anomalies over the Niño 3.4 region in the equatorial Pacific Ocean. The corresponding time series, shown in Fig. \ref{fig:datasets}(b), is available for download from \cite{NOAA_Nino34_web}. A detailed description of the underlying data set and methodology is provided in \cite{Rayner2003}.} We use this time series as a paradigmatic case of a stationary time series  where the relevant events appear as local extreme values.

{The third series analyzed is}
{the Global Mean Land/Ocean Temperature (GMT) Anomaly Index, which is an estimate of global surface temperature change. This data is provided by NASA's Goddard Institute for Space Studies \cite{GISTEMPv4_2026}. 
The time series is shown in Fig. \ref{fig:datasets}(c) and it is available for download from \cite{NOAA_gmsst_web}. More information concerning methodology and analysis of this dataset can be found at \cite{Lenssen2024_noaa}.}  We use this time series as a paradigmatic case of a nonstationary time series, where the source of nonstationarity resides in the first moment of the marginal distribution (increasing mean).

{The forth time series analyzed, }
{shown in Fig. \ref{fig:datasets}(d), is an unbiased random walk \cite{shumway2011timeseries}, defined by $x_{t+1} = x_t+\xi$, in which $\xi$ is an uncorrelated Gaussian noise. We use this final example as a nonstationary stochastic process --whose source of nonstationarity is located in the second moment in terms of the ensemble--, where the direct identification of extreme values in the time series is not trivial}.



\section{Method}\label{sec:methods}

{Let us start by introducing the VG method \cite{Lacasa2008}, which maps a time-ordered sequence $Y=[(t_1,y_1),(t_2,y_2),...(t_n,y_n)]$ of $n$ data points into a labeled graph of $n$ nodes, where node $k$ is related to the $k$-th data point and two nodes $a$ and $b$ are connected in the VG if their associated data values
 $(t_a,y_a)$ and $(t_b, y_b)$ are such that any intermediate data $(t_c, y_c)$ such that $t_a<t_c<t_b$ satisfy}
\begin{equation}
y_c < y_b + (y_a - y_b) \frac{(t_b - t_c)}{(t_b - t_a)} , \quad \forall c: t_a<t_c<t_b.
\label{eq:vg}
\end{equation}

{The resulting graph inherits the structure of the time series in its topological structure.}

In 
this work, we will also generate the VG of the time series subsample containing only the so-called peak values, i.e. those points in the time series that are local maxima.
Specifically, 
a data point at time $t_i$ is considered a peak if $y_{i+1}<y_i$ and $y_i>y_{i-1}$. The so-called {\it peak time series} is then sampled from the original time series. In general the peak time series is not equispaced, and it is easy to see that the associated VG is a subgraph of the VG associated to the original time series. 

\begin{figure}[tb] 
\centering
\includegraphics[width=0.95\linewidth]{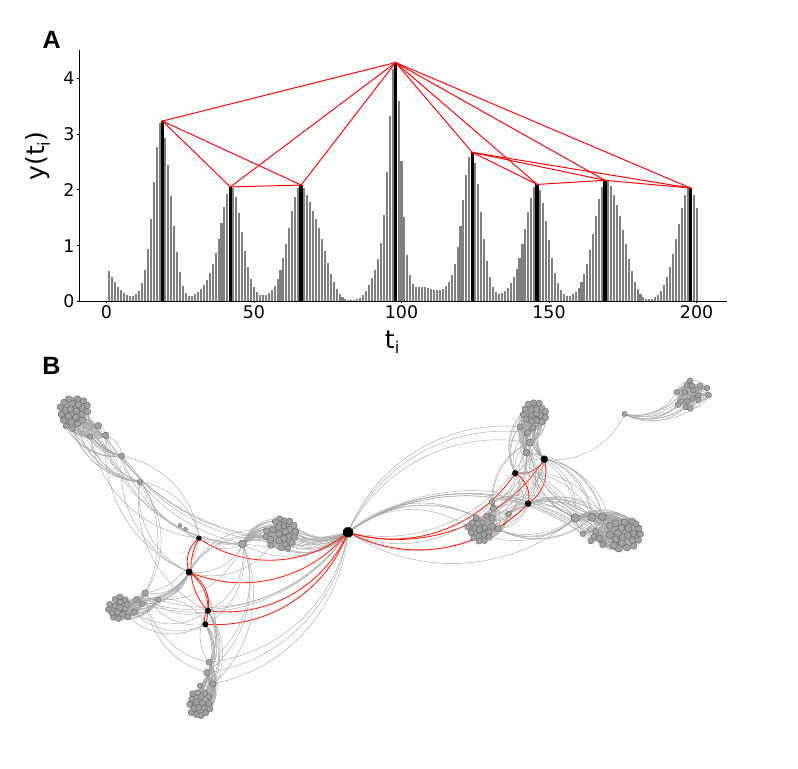} 
\caption{\justifying (a) Example of a time series and (b) visibility graph (VG) constructed from the peak time series shown in (a) (black bars). Red lines in (a) illustrate the visibility criterion of Eq. (\ref{eq:vg}) and correspond to the red edges in the VG in (b). }
\label{fig:vg_subsamp}
\end{figure}

The VG method and the sub-sampling are illustrated in Fig.~\ref{fig:vg_subsamp}. In panel (a), the original time series (from the deterministic rogue wave model described in Sec.~\ref{sec:data}) is shown in gray bars. Also in this panel, we show the peak values in black bars, as well as the connections of the VG of the resulting peak time series in red lines. Panel (b) displays the VG from the original time series in gray and the VG from the peak time series with connections in red and nodes in black. Using a sub-sample of the time series can significantly decrease computational time for the VG algorithm, and we show in the next section that some data values, that are locally extreme,  can also be detected employing this approach.

{After obtaining the VG associated to the time series, }
{multiple analyses can be performed to characterize {the graph, i.e., to obtain graph-based features that encapsulate properties of the individual data values (the nodes in the VG) and of the global} time series. In particular, 
{metrics that quantify the connectivity of the nodes --so-called centrality metrics such as degree, betweeness or subgraph centrality \cite{Boccaletti_2006, estrada2012structure, latora2017complex}  will give information on the local importance of nodes in the graph, and so, of the corresponding data points in the time series.  
{In this work we use the simplest centrality metric:} the node's degree, measuring its number of connections/edges in the VG. The rationale for using this metric is based on previous theoretical works \cite{luque2009horizontal, lacasa2018visibility} that found quantitative relations between the $t$-th time series value $x$ and the degree $k$ of its associated $t$-th node. In particular, in the {horizontal visibility graph (HVG) \cite{luque2009horizontal}, which provides a subgraph of the VG one, Luque {\it et. al} rigorously proved \cite{luque2009horizontal}} that the ensemble-averaged degree associated to a data value $x$ follows, when the time series is uncorrelated and has marginal distribution $f(x)$
    \begin{equation}
        k(x)=2-2\ln[1-F(x)],
        \label{eq:theory}
    \end{equation} where $F(x)$ is the cumulative distribution function (CDF) of $f(x)$  \cite{luque2009horizontal}. {This relation 
    was found to hold, on average, 
    also for chaotic processes with weak or vanishing correlations \cite{lacasa2018visibility}.}

We note that the ensemble-average can be traded for a long single realization of the time series by performing binning averages. In the Appendix 
we illustrate the process of binning average and show how the theoretical prediction is exact for HVGs associated to uncorrelated time series, and holds to a good approximation also in the VG for a range of stationary processes with different correlations. While VG is less amenable to analytical insight that the HVG, it is nonetheless more expressive --particularly in nonstationary processes having long-range memory \cite{lacasa2009visibility}-- and is therefore the method of choice in this work. Our numerical analysis in a range of synthetic time series shows that Eq.~(\ref{eq:theory}) remains a reasonably good approximation to $k(x)$ for VG of uncorrelated processes and stationary processes with correlations.

The key conclusion of this theoretical insight is that, for stationary processes, on average there is (i) monotonic and (ii) nonlinear relation between the value of a point in the time series and its degree. Monotonicity guarantees that high data values relate on average to nodes with high degree nodes, and justifies the use of ranked data, as we will use below. Importantly, the nonlinear amplification of Eq.~(\ref{eq:theory}) enhances the contribution of large values while suppressing noise. In other words,  whereas for small data values the relation between data value and node degree holds on average, for the extreme values of the time series such relation will hold without the needs to perform any average. Finally, Eq.~(\ref{eq:theory}) completely breaks down for nonstationary processes. This is expected, since for nonstationary time series the value of a data point is not really an indication of its character as extreme event, however, in what follows we will show that the node's degree centrality will remain as a useful indicator in this more challenging scenario.

\section{Results} \label{sec:results}
\subsection{Stationary time series}
{We begin by illustrating the method using the ORW time series. After generating the corresponding VG, we rank and order both the data point values (aka heights) in the time series and their associated node degrees in the VG (see
Tables~\ref{tab:ranks}(a) and \ref{tab:ranks}(b) for an illustration of the first six positions of both rankings, alongside time index information).
It can be observed that at the top of both rankings we have the same time index (i.e. the datum at time index 1201 has the heighest value in the time series and also the largest degree in the associated VG), but this relationship does not strictly hold in general. 
For example, the data points with indices 553 and 930 have the same degree, so they occupy the same position in the rank by degree, and neither of them appears in the first six positions in the rank by height.}


\begin{table}[tb]
\centering
\caption{VG ranking method for the ORW time series shown in Fig. \ref{fig:ts_rw}. The first six positions in the ranks by height of the time series values and degree of nodes in the VG are displayed. }
\begin{subtable}{0.48\textwidth}
\centering
\vspace{-0.3cm}
\caption{Ordering by height}
{
\renewcommand{\arraystretch}{1.3}
\begin{tabular}{>{\columncolor{lightgray}}c|c|c|c|c}
\Xhline{1.2pt}
rank by height & rank by degree & index & height & degree \\
\Xhline{1.2pt}
1 & 1 & \textbf{1201} & 10.5 & 221 \\
\hline
2 & 4 & 1200 & 8.3 & 104 \\
\hline
3 & 6 & 1202 & 7.7 & 77 \\
\hline
4 & 2 & 298  & 6.5 & 138 \\
\hline
5 & 3 & 1378 & 6.1 & 119 \\
\hline
6 & 7 & 297  & 6.0 & 71 \\
\Xhline{1.2pt}
\end{tabular}
}
\end{subtable}
\hfill
\vspace{0.1cm}
\begin{subtable}{0.48\textwidth}
\centering
\caption{Ordering by degree}
{
\renewcommand{\arraystretch}{1.3}
\begin{tabular}{c|>{\columncolor{lightgray}}c|c|c|c}
\Xhline{1.2pt}
rank by height & rank by degree & index & height & degree \\
\Xhline{1.2pt}
1  & 1 & \textbf{1201} & 10.5 & 221 \\
\hline
4  & 2 & 298  & 6.5  & 138 \\
\hline
5  & 3 & 1378 & 6.1  & 119 \\
\hline
2  & 4 & 1200 & 8.3  & 104 \\
\hline
27 & 5 & 553  & 4.0  & 88 \\
\hline
10 & 5 & 930  & 4.7  & 88 \\
\hline
3  & 6 & 1202 & 7.7  & 77 \\
\Xhline{1.2pt}
\end{tabular}
}
\end{subtable}
\label{tab:ranks}
\end{table}

{These top-ranked values listed in Tables~\ref{tab:ranks}(a) and \ref{tab:ranks}(b) are also highlighted}
{depicted in Fig. \ref{fig:ts_rw}. In panel (a) of this figure, we show that the points identified in the rank by height, which could also be found by using an appropriate} thresholding. On the other hand, in panel (b), some 
not-too-extreme data points (subthreshold) are identified in the rank by degree, pointing that these `subthreshold' values are possibly important in relation to their neighbors in the time series. In summary, we can   identify EVs both by height directly on the time series but also by degree.}

\begin{figure*}[tb] 
\centering
\includegraphics[width=\linewidth]{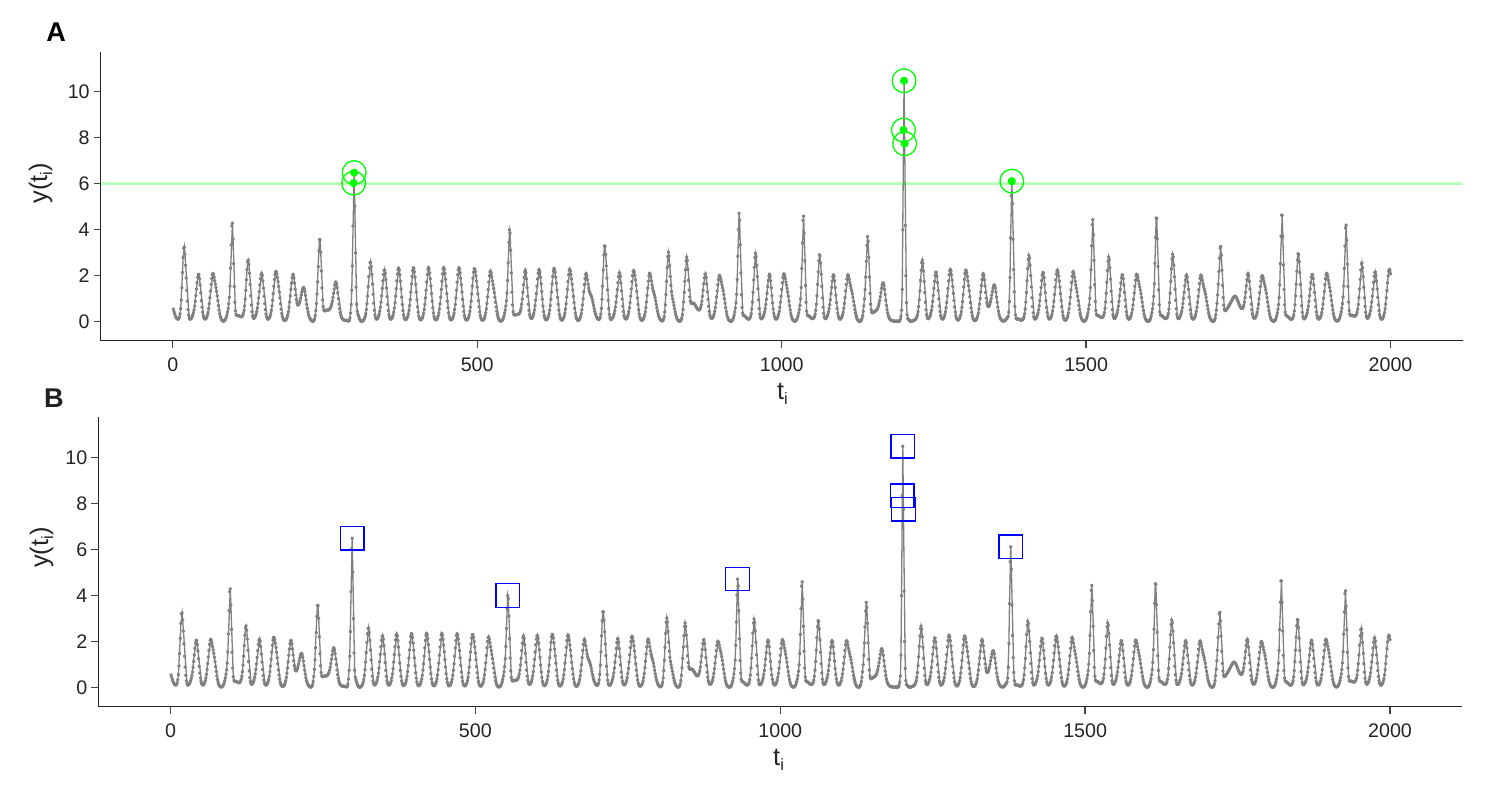} 
\caption{\justifying Illustration of extreme values (EVs) detected by height and by degree, in the laser intensity 
time series shown in Fig. 1(a). 
Panels (a) and (b) show the EVs identified by the height (open green circles) and by the degree (open blue squares) of the nodes in the generated visibility graph, respectively. These values are also listed in Table~\ref{tab:ranks}. Solid green circles in panel (a) are the values above the threshold depicted by the green line. }
\label{fig:ts_rw}
\end{figure*}

{Figure \ref{fig:kvsh_ranks}(b) displays all rank positions, ordered by height and degree, for the aforementioned ORW time series. Each point in this panel corresponds to a node of the VG (and to an index of the data points), and the nodes shown in the inset have already been identified in Tab. \ref{tab:ranks}. Analyzing this point cloud, it can be observed that the majority of points are concentrated near the upper region of the plot, while the nodes classified as having high degree/height converge toward the plot origin. This distinct separation pattern occurs for stationary time series containing pronounced EVs. In contrast, an analogous behavior is not discernible when employing a direct relationship between degree and height, as illustrated in Fig. \ref{fig:kvsh_ranks}(a).}

\begin{figure}[tb] 
\centering
\includegraphics[width=\linewidth]{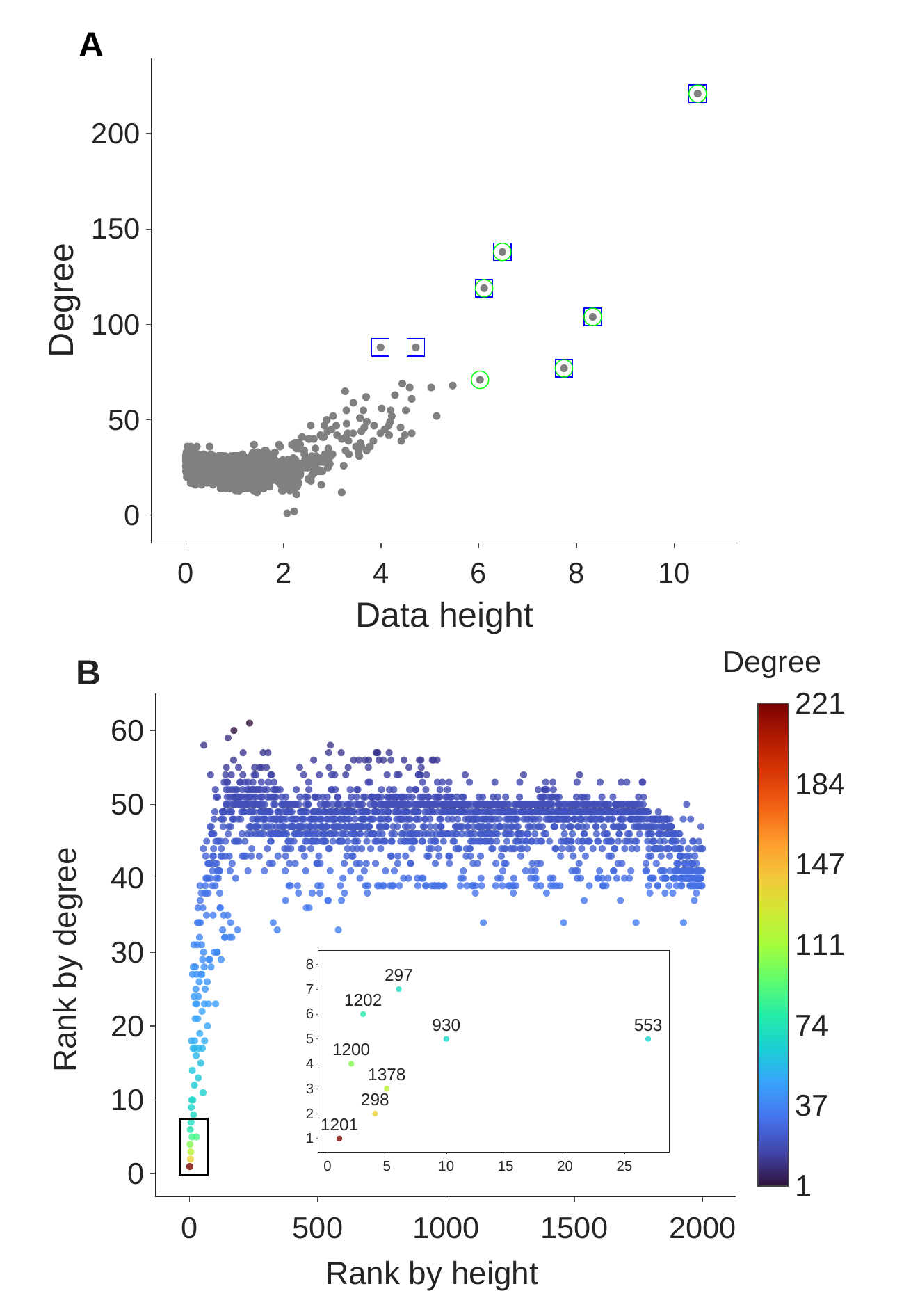} 
\caption{\justifying Detection of extreme values using the visibility graph. Panel (a) shows the node degree in the VG as a function of the corresponding height in the time series shown in Fig. \ref{fig:ts_rw}. Open blue circles and green squares indicate the events identified by the degree and height criteria, respectively, in the same figure. Panel (b) presents the ranking by height and degree obtained from the VG of the same ORW time series. The highlighted region is zoomed in to explicitly show the indices of the EVs identified.}
\label{fig:kvsh_ranks}
\end{figure}


{In the following, we characterize the EV present in the first two time series described in Sec. \ref{sec:data}, starting with the ORW time series. As already shown in Fig. \ref{fig:kvsh_ranks}(b), the nodes displayed in the inset correspond to the EV identified by selecting the top six ranked nodes in both height and degree, when the visibility graph is constructed from the complete ORW time series. When the analysis is restricted to a sub-series comprising only of peak values, a very similar point cloud is obtained for the ranks of degree and height of the peak-based visibility graph, $\text{VG}_\text{peaks}$, as illustrated in Fig. \ref{fig:rank_full_peaks_VG}(a). From these ranks, we again select the six highest-ranked nodes to identify the corresponding EV.}
{These EV are associated with the nodes shown in Fig. \ref{fig:rank_full_peaks_VG}(b), and most of them can be directly related to the EV previously identified from the VG built on the full time series. It is important to note that, when using $\text{VG}_\text{full}$, multiple points belonging to the same EE will be selected simultaneously in the ranking, both by height and degree, depending on the temporal resolution of the data. For instance, the data point with index 1201 in Fig. \ref{fig:ts_rw} forms, together with its neighboring points, a group that is associated with a single EE in the system dynamics. This redundancy is mitigated when the time series is pre-processed such that only the peaks are used.}

{The EVs identified in Fig. \ref{fig:rank_full_peaks_VG}(b) are also highlighted in the time series shown in panel (d). In this representation, additional EVs become apparent because the ranking changes when only peak values are considered. The VGs constructed from the full time series and from the peak-filtered series are depicted in Fig. \ref{fig:rank_full_peaks_VG}(c) in gray and black, respectively. In this panel, node size encodes the degree, while the node color encodes the EV classification corresponding to panels (b) and (d).}

\begin{figure*}[tb] 
\centering
\includegraphics[width=\linewidth]{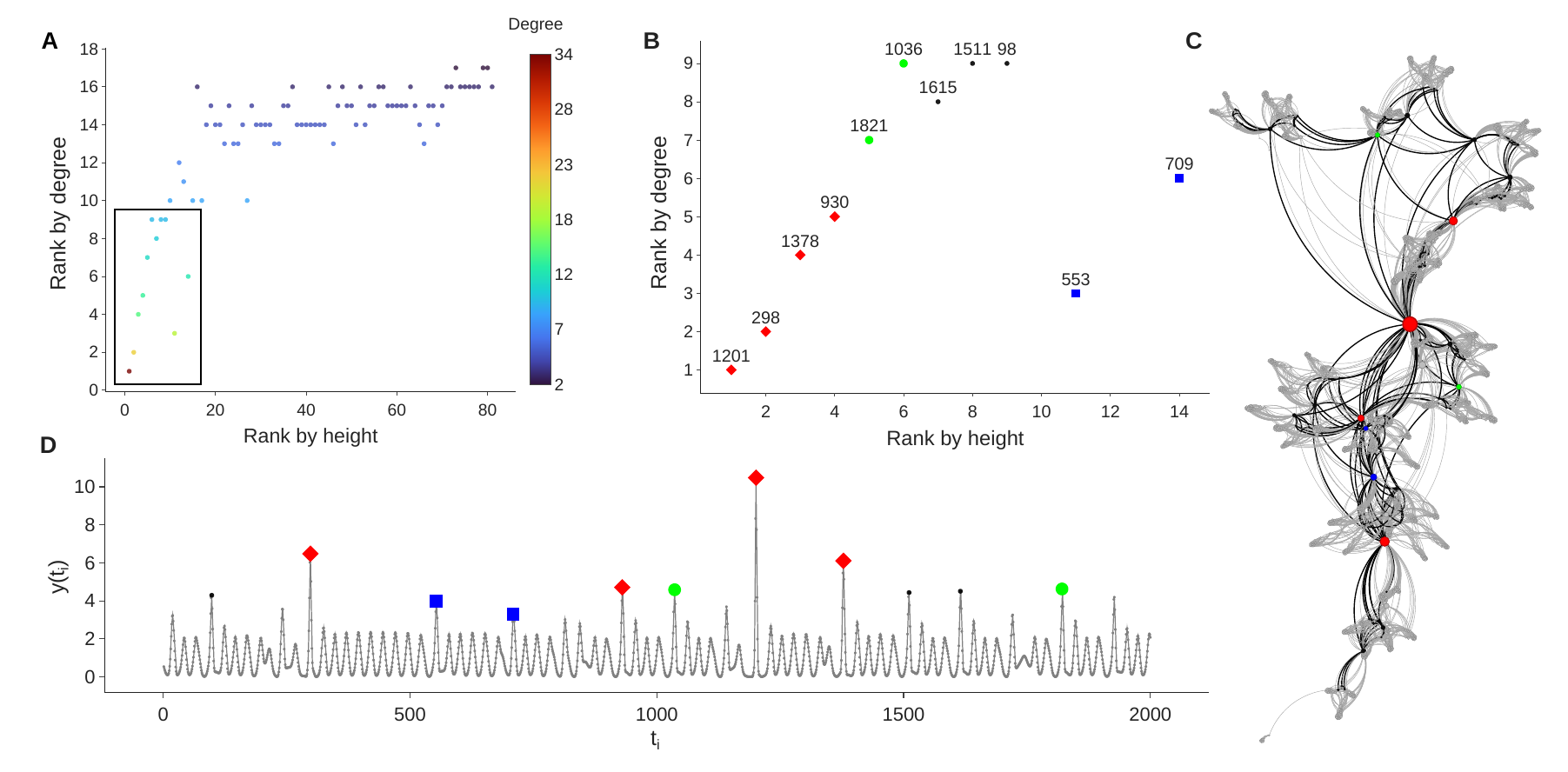} 
\caption{\justifying Visibility graph ranking method applied to the laser intensity time series, shown in Fig. 1(a). 
Panel (a) shows the ranking by degree and height. The highlighted region in (a) is zoomed in panel (b), where the marked points by green circles, blue squares, and red diamonds are the EV identified by the degree, the height, and by both criteria, when the first six positions in both ranks are considered. These points are also marked in the time series shown in panels (d) and (c). Panel (c) displays the VG constructed from the full time series in gray and from the peaks in black.}
\label{fig:rank_full_peaks_VG}
\end{figure*}

{Incidentally, observe that the number of positions kept in the rankings is a free parameter which will determine the number of EV that are identified. So that the correct procedure is to increase the positions analyzed one by one and observe which points are identified, but always using the same number of positions in both ranks.}

The second case we address is the Niño3.4 SST Anomaly Index time series, which is again a stationary one. The rank plot for this data is shown in Fig. \ref{fig:ranks_nino3.4}(a) and (b), for the $\text{VG}_\text{full}$ and $\text{VG}_\text{peaks}$, respectively. It is easy to notice that the shape of the point cloud is preserved after the sub-sampling. We can again delimit the top six positions in the ranks, so that the inset of Fig. \ref{fig:ranks_nino3.4}(b) can be zoomed in Fig. \ref{fig:ranks_nino3.4}(c) to show the identified EVs. These EVs are also marked in the time series in Fig. \ref{fig:ranks_nino3.4}(d). 

These EVs tend to correspond to El Niño episodes; however, an important observation is that the data points with indices 720 and 635 are only identified by degree, and they correspond to moderate El Niño-like episodes. Therefore, standard methods considering only the amplitudes (heights) in the time series would fail to identify some EVs with local importance, which are captured by the VG nodes' degree. Additionally, with this result, we show that the method successfully identifies the EVs even if they are not very pronounced, as in the first case of the ORW time series.



\begin{figure*}[tb] 
\centering
\includegraphics[width=\linewidth]{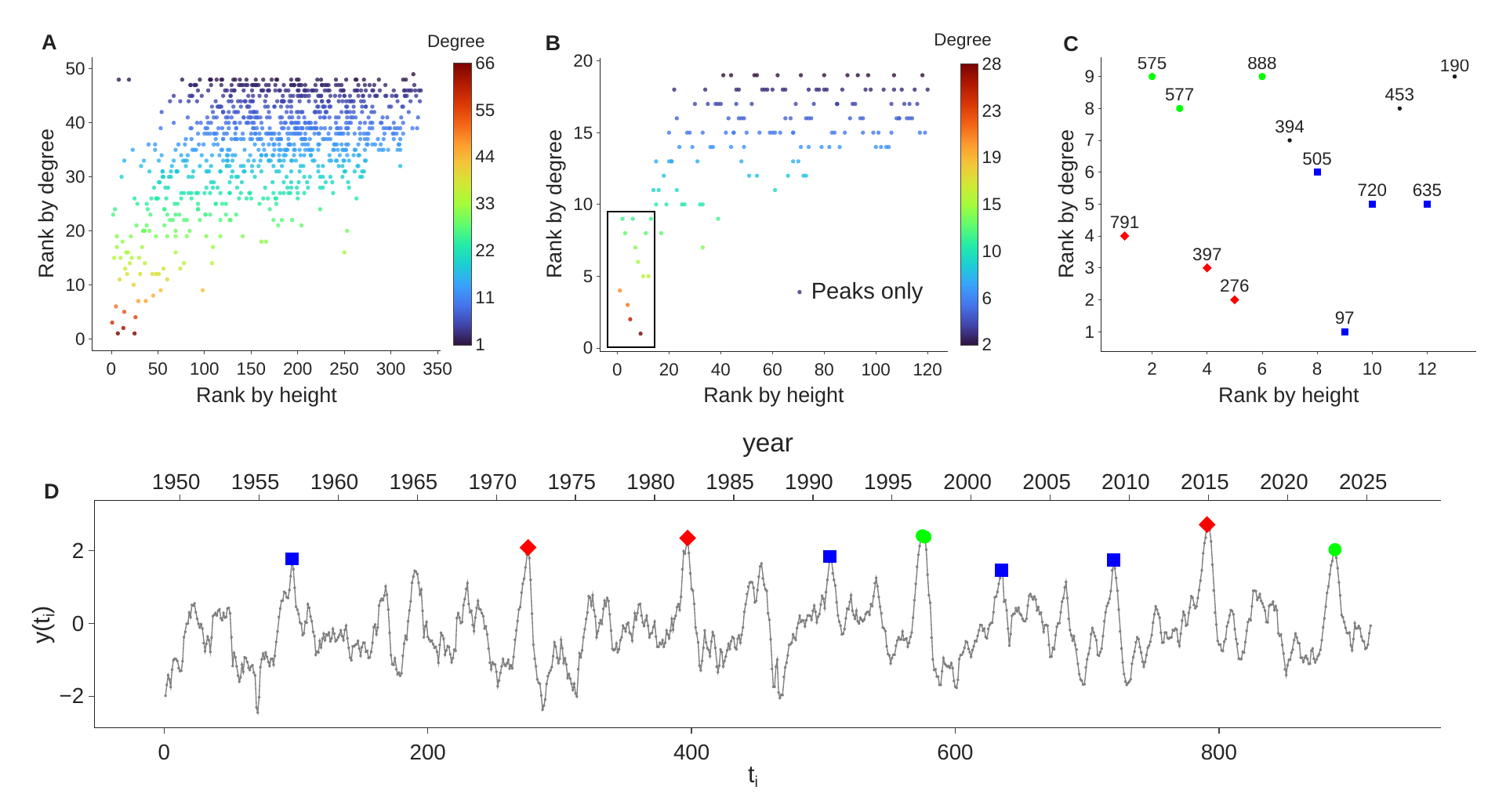} 
\caption{\justifying VG ranking method applied to the time series shown in Fig. 1(b), el Niño3.4 SST Anomaly Index \cite{NOAA_Nino34_web}. Panels (a) and (b) show the ranks by height and degree for the full time series and the peak time series, respectively. The highlighted region in (b) is zoomed in panel (c), where points marked in green circles, blue squares, and red diamonds are the EV identified by the degree, the height, and by both criteria, respectively,  when the first six positions in both ranks are considered. These EV are also marked in the time series in panel (d) with the same code color.} 
\label{fig:ranks_nino3.4}
\end{figure*}


\subsection{Nonstationary time series}
As 
explained in the Introduction, the traditional concept of an extreme event is intimately related to the idea of data outliers: points in the time series that are extremely different from the rest. Quantification of outliers --of which extreme events are a particular type-- boils down to (i) estimate the marginal distribution $f(x)$ of the time series $\{x_t\}$, (ii) assess for which time points $t$, we have that $x_t$ is `not in $f(x)$'. Observe that this concept is thus well-defined only when $f(x)$ itself is well-defined. This is the case for stationary processes, for which $f(x)$ does not change over time. Accordingly, analysis of extreme events in nonstationary processes --whose data marginal distribution varies over time-- is often considered anathema.

Let us consider, for instance, an unbiased random walk $x_{t+1}=x_t+\xi$, where $\xi\sim g(x)$ is an uncorrelated noise with marginal distribution $g$. It is well known that the marginal distribution of the random walk changes over time as its variance (related to the second moment of the distribution) grows linearly with time. Accordingly, for random walk time series it is not possible to unambiguously define what an extreme event is.

Interestingly, when the time series is mapped into a visibility graph, it is well known that the VG is invariant under certain transformations of the signal, including linear trends \cite{Lacasa2008}. Moreover, the degree distribution (i.e. the marginal distribution of the degree sequence) is indeed a stable property across a range of dynamical processes which, while nonstationary, are VG-stationary \cite{lacasa2015time}, like unbiased random walks with our without superimposed linear trends.
In \cite{lacasa2009visibility} it was indeed found that fractional Brownian motion with Hurst exponent $H$ (of which the case $H=1/2$ generates the special case of Brownian motion, i.e. random walk time series) are nonstationary processes (with time-varying marginal distributions) whose visibility graphs are VG-stationary and have a degree distribution which does not change over time $P(k)\sim k^{2H-3}$. In other words, this indicates that while extreme events in e.g. time series with superimposed trends, random walks or  fractional Brownian motion cannot be directly defined in terms of data height rankings, the definition of these extremes in terms of degree rankings is still perfectly possible.

{To illustrate the application of our method to nonstationary processes, we analyze both real and synthetic generated data. The first case is the time series of the Global Mean Land/Ocean Temperature Anomaly Index from 1948 to 2026. The ranking plot is shown in Fig. \ref{fig:ranks_temp_global}(a). Although the ranks of degree and height do not show a well-separated region with the EVs, it can be observed in Fig.~\ref{fig:ranks_temp_global}(b) that the nodes occupying the first positions in the rank by degree distribute across the time series, and are related to locally important events or local temperature records. This suggests that it is possible to identify local EVs using the degree of the nodes in the VG for nonstationary time series, without the needs to detrend or stationarise the signal, and without the needs of pre-defining local windows of observation, something which in principle is challenging and not principled. Finally, a similar behavior happens when analyzing a time series from an unbiased random walk. As shown in Fig. \ref{fig:ranks_random_walk}(b), the data points identified only by height concentrate at the beginning of the descending curve, while the ones identified by degree distribute across the time series. These identified nodes can also be found in the ranking plot of Fig. \ref{fig:ranks_random_walk}(a). 


\begin{figure*}[tb] 
\centering
\includegraphics[width=0.95\linewidth]{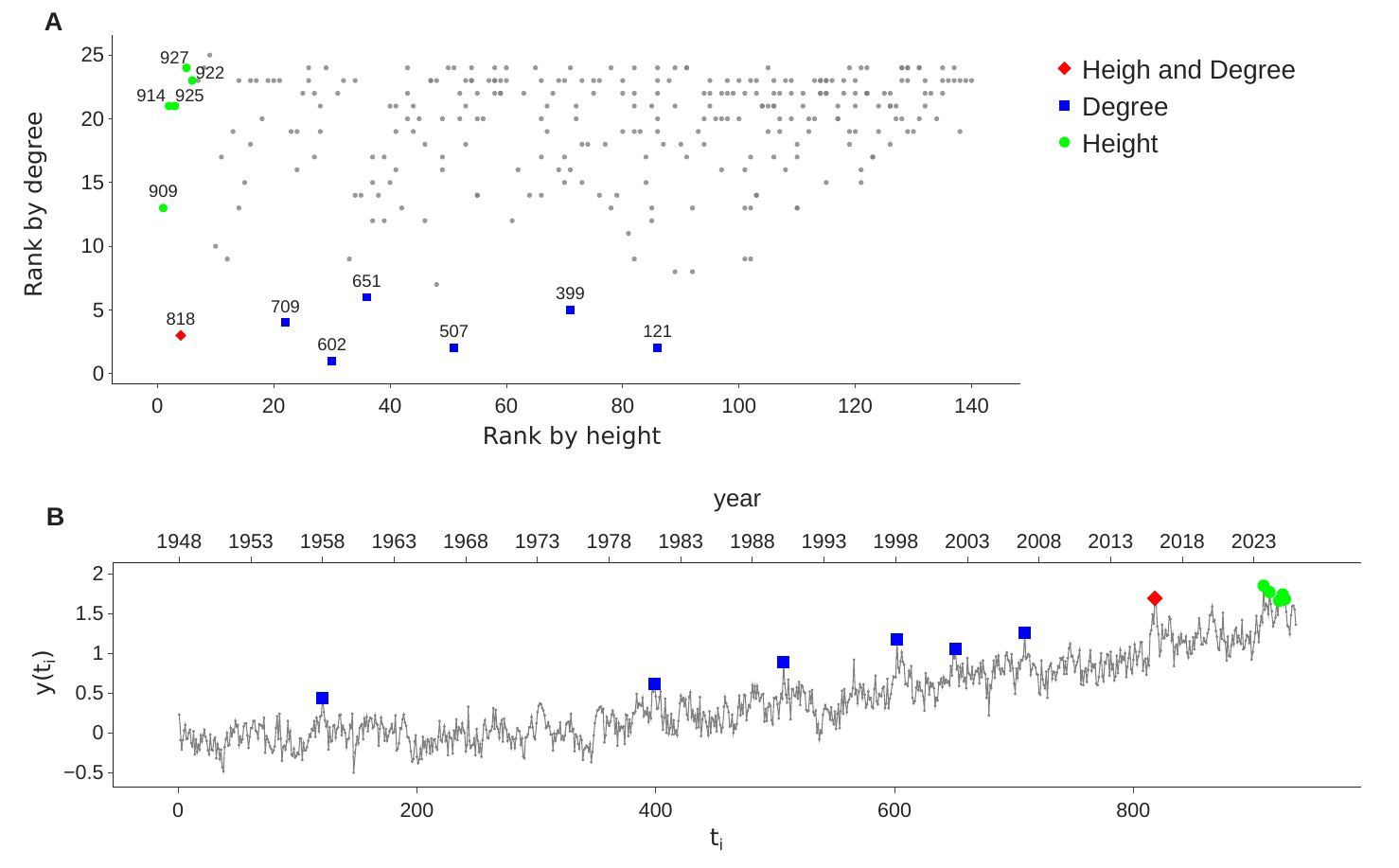} 
\caption{\justifying VG ranking method applied to the time series shown in Fig. 1(c), the Global Mean Land/Ocean Temperature Anomaly Index \cite{NOAA_gmsst_web}. Panel (a) shows the ranks by height and degree for the VG generated from the peak time series. The marked points in green circles, blue squares, and red diamonds are the EV identified by height, degree, and by both criteria, respectively, when the first six positions of both ranks are considered. These points are also marked in the time series in panel (b) using the same code color.} 
\label{fig:ranks_temp_global}
\end{figure*}

\begin{figure*}[tb] 
\centering
\includegraphics[width=0.95\linewidth]{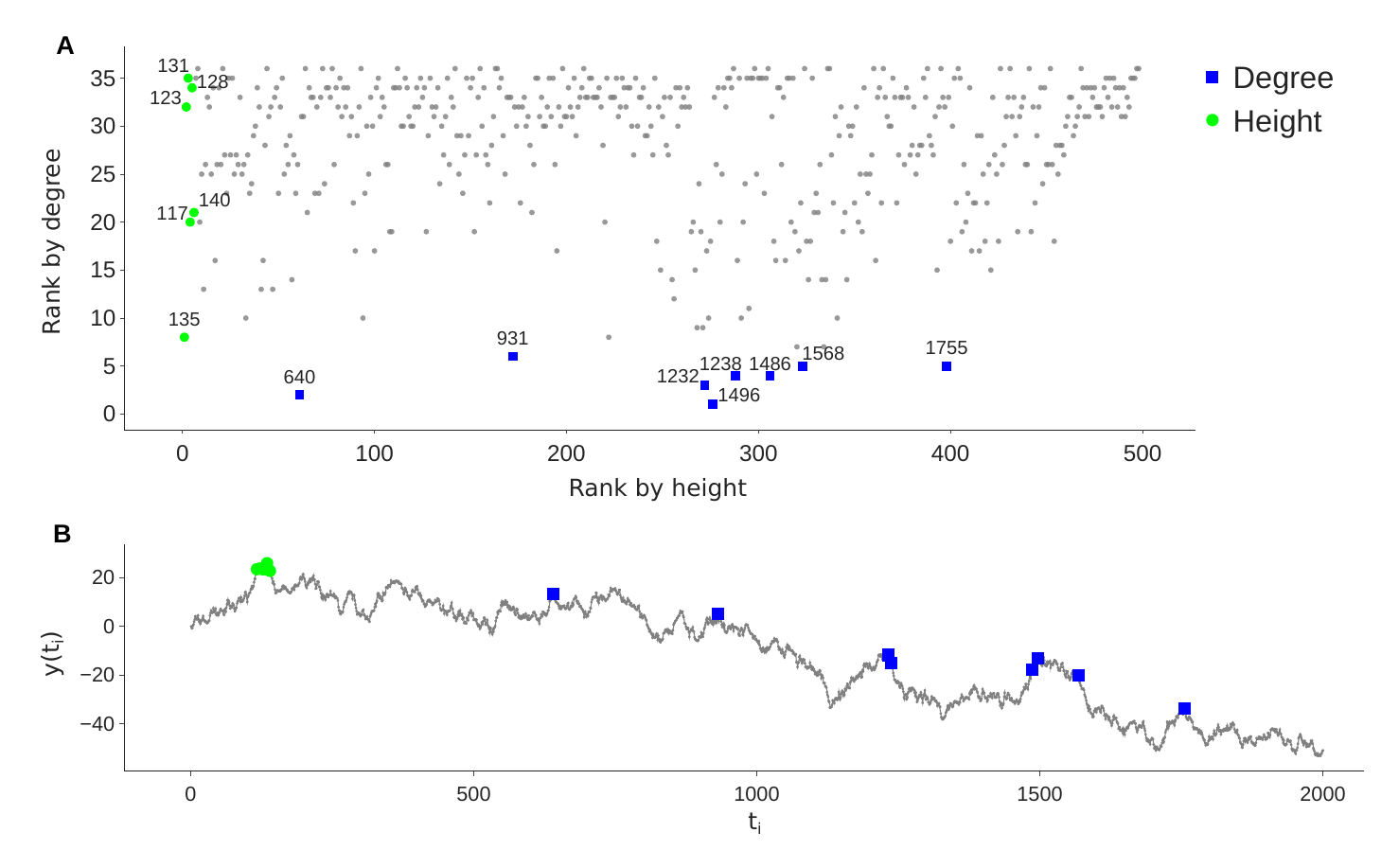} 
\caption{\justifying VG ranking method applied to the time series shown in Fig. 1(d), generated by an unbiased random walk. Panel (a) shows the ranks by height and degree for the VG generated from the peak time series. The marked points in green circles and blue squares are the EV identified by height and degree, respectively, when the first six positions in both ranks are considered. These points are also marked in the random walk time series in panel (b) using the same code color. } 
\label{fig:ranks_random_walk}
\end{figure*}

\section{Discussion and Conclusions} \label{sec:conclusion}

{In this work, we 
propose a methodology for the detection and characterization of extreme events. Specifically, we employ a mapping of the time series onto a visibility graph, in which each data point is represented as a node in a complex network. Within this framework, we propose an approach based on the degree centrality of the nodes. By jointly ranking all data points according to their amplitude and their degree in the visibility graph, we demonstrate that extreme values systematically occupy the top positions in these rankings. Also, we show that, in general, it is not necessary to analyze the complete time series. A reduced subset of data points consisting solely of local maxima (peaks) is often sufficient to correctly identify the extreme values, reducing the computational cost while preserving detection accuracy.}

{Furthermore, the advantage of employing the degree-based ranking instead of only the height-based ranking lies in the fact that a node’s connectivity is intrinsically linked to its relative importance with respect to its neighbors. This allows for the identification of locally extreme values that would be overlooked when using only a global threshold criterion.}

To demonstrate the methodology, we first use a time series obtained by numerical simulation of a well-established model of semiconductor lasers, in which the extreme events manifest as extreme pulses also known as optical rogue waves. However, the method remains applicable even when the extreme values are less pronounced or exhibit less distinct shapes. This is exemplified by the analysis of Niño3.4 index, 
where the rankings not only capture the most prominent extremes in the time series but also detect local extremes associated with moderate El Niño-like} episodes.

The proposed method 
is particularly advantageous for the analysis of nonstationary time series, such as the Global Mean Land/Ocean Temperature Anomaly. For this time series, the conventional threshold-based criterion is not appropriate, nor is the rank by height, which, due to global warming, will identify data values concentrated in the last few years. 
Under these conditions, the identification of extreme events based on visibility degree 
becomes especially informative and thus even more valuable for characterizing the system of interest.

For future work, it will be interesting to analyze 
how the VG-based identification of extreme values relates to system-specific events such as regime shifts or the onset of exogenous events. Finally, applications of this methodology extend beyond the cases considered in this work, including rainfall, financial time series, neural activity 
and physiological data (e.g. heartbeat signals), to mention just a few examples.



\section*{Appendix: Relation between degree $k$ and time series data value $x$.}

In Fig.~\ref{fig:uniform_WN} we show the scatter plot of $k$ vs $x$ extracted from the VG or HVG, when time series are generated via different dynamics. Panels (A,B) display the results of HVG and VG, respectively, associated to a white (uncorrelated) time series with uniform  marginal distribution. We average out fluctuations by binning data (in all panels, we set 100 equispaced bins over the whole range of $x$ values). The dashed line is the theoretical Eq.~\ref{eq:theory}, which shows good agreement with the binned average for the HVG (Fig.~\ref{fig:uniform_WN}A), and is only an approximation for the VG (Fig.~\ref{fig:uniform_WN}B). What is important is that the monotonic behavior is preserved, with an amplification towards higher $x$ that conveys $k$ good properties as extreme event quantifier.

We note that Eq.~\ref{eq:theory} is monotonic for all $F(x)$, since $F$ is a CDF and thus monotonically increasing. Monotonicity already implies $\text{rank}(k)=\text{rank}(x)$. Now, such relation is only on average, since Eq.~\ref{eq:theory} provides an average relation. However, the particular shape in Eq.~\ref{eq:theory} amplifies the degrees of very large outliers, observe the logarithmic divergence for outliers located at the right-end of the CDF. This suggests that the relation $\text{rank}(k)=\text{rank}(x)$ is likely to hold, even in finite time series, for the points of very high $x$, due to the logarithmic divergence amplification in $k(x)$.

Figure~\ref{fig:uniform_WN}(C) shows the results for an AR(1) process $x_{t+1}=0.8x_t + \xi$, $\xi\sim N(0,1)$. This process is stationary and weakly correlated, i.e. autocorrelations decay exponentially fast. The tendency and motononicity holds although the theoretical prediction is less accurate, as expected.

Now, for correlated time series, Eq.~\ref{eq:theory} is no longer necessarily valid, but the monotonic trend remains. This holds also for long-range correlated stationary processes, see Fig.~\ref{fig:uniform_WN}(D) where a fractional Gaussian noise (fGN) is analysed. fGN is a stationary process with long-range autocorrelation (power law with an exponent which is linear in $H$, the Hurst exponent).

Finally, an even more dramatic situation is nonstationary processes, like random walks or fractional Brownian motion, time series that emerge in realms like geoscience (temperature records, rainfall, etc), finance, neuroscience, physiology (heartbeats), etc. In this case, extreme events cannot be characterised in terms of outlier values of the marginal distribution $f(x)$, since the latter changes over time. In particular, for Brownian motion (random walks) $\sigma \sim t^{1/2}$, and for fBm $\sigma \sim t^{H}$. So an outlier in the first $N$ points of a time series won't be classified as an outlier anymore for the first $2N$ points of the series because the standard deviation of the signal increased and so the threshold to define what an outlier is.
However, in these cases, the node's degree $k$ is still well-defined, i.e. the marginal degree distribution $P(k)$ --unlike $f(x)$-- does not change its shape for nonstationary processes \cite{lacasa2009visibility}. At the same time, in the case of a random walk, shown in Fig.~\ref{fig:uniform_WN}(E), we can see that there is no relation really between $k$ and $x$. This 
suggests that $k$ or $\text{rank}(k)$ could be used here for extreme event detection (the hubs) in a situation where extreme events cannot be detected by $x$.

\begin{figure*}[htb!]
\centering
\includegraphics[width=\linewidth]{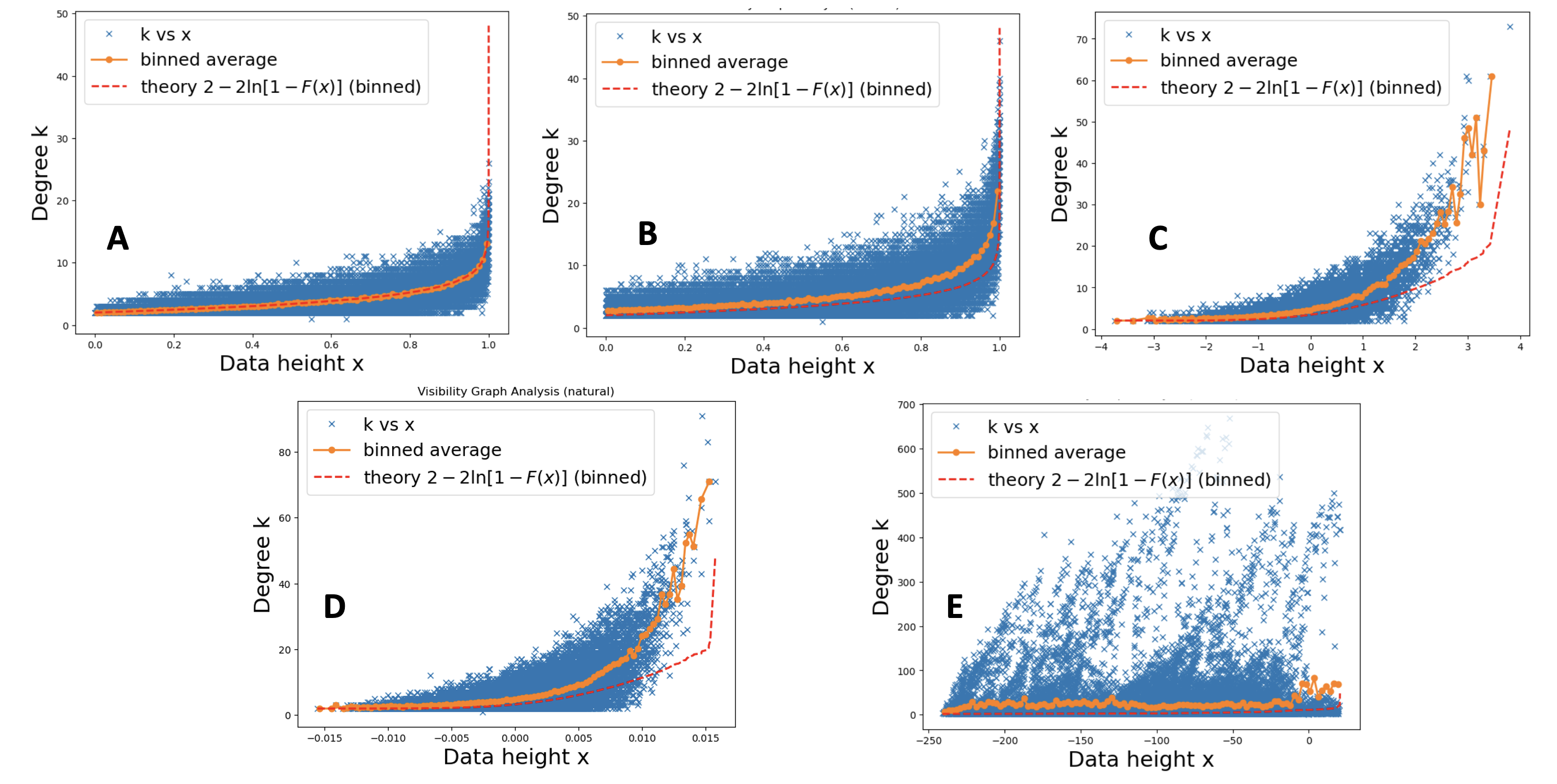} 
\caption{Degree $k$ of a node vs its associated time series data value $x$. In all panels, time series have $N=3\times 10^4$ points. In each panel, the VG/HVG is generated from a time series with different dynamics. In all panels, blue crosses correspond to the $k$ vs $x$ scatter plot, orange dots correspond to the binned averages (binning the range of $x$ values in 100 equispaced bins), and the red dashed line is Eq.~\ref{eq:theory}, which provides the ensemble-averaged relationship, only exact for uncorrelated time series under the HVG transformation, and otherwise being an approximation. 
(A) White uniform noise, HVG. (B) White uniform noise, VG. (C) Autorregressive process $x_{t+1}=0.8x_t + \xi$, $\xi \sim N(0,1)$, VG. (D) Fractional Gaussian noise with Hurst exponent $H=0.7$, VG. (E) Unbiased Gaussian random walk $x_{t+1}=x_t + \xi, \xi\sim N(0,1)$.}
\label{fig:uniform_WN}
\end{figure*}

\section*{Acknowledgments} J. T. M. and C. M. acknowledge support from project PID2024-160573NB-I00 funded by Agencia Estatal de Investigacion (AEI), Spain. L. L. acknowledges partial support from project CSxAI (PID2024-157526NB-I00) funded by 

MICIU/AEI/10.13039/501100011033/FEDER, UE, from a Maria de Maeztu grant (CEX2021-001164-M) funded by

the MICIU/AEI/10.13039/501100011033, and from the European Commission Chips Joint Undertaking project No. 101194363 (NEHIL).

\bibliographystyle{elsarticle-num}
\bibliography{refs}

\begin{thebibliography}{10}
\expandafter\ifx\csname url\endcsname\relax
  \def\url#1{\texttt{#1}}\fi
\expandafter\ifx\csname urlprefix\endcsname\relax\def\urlprefix{URL }\fi
\expandafter\ifx\csname href\endcsname\relax
  \def\href#1#2{#2} \def\path#1{#1}\fi

\bibitem{book_kantz}
S.~Albeverio, V.~Jentsch, E.~Kantz, H., Extreme Events in Nature and Society, Springer, 2006.

\bibitem{beggs2003neuronal}
J.~M. Beggs, D.~Plenz, Neuronal avalanches in neocortical circuits, Journal of neuroscience 23~(35) (2003) 11167--11177.

\bibitem{sornette2003critical}
D.~Sornette, Critical market crashes, Physics reports 378~(1) (2003) 1--98.

\bibitem{book_rws}
C.~Kharif, E.~Pelinovsky, A.~Slunyaev, Rogue Waves in the Ocean, Springer, 2009.

\bibitem{solli_2007}
D.~R. Solli, C.~Ropers, P.~Koonath, B.~Jalali, Optical rogue waves, Nature 450 (2007) 1054.

\bibitem{rw_prl_2011}
C.~Bonatto, M.~Feyereisen, S.~Barland, M.~Giudici, C.~Masoller, J.~R. Rios~Leite, J.~R. Tredicce, Deterministic optical rogue waves, Phys. Rev. Lett. 107 (2011) 053901.

\bibitem{kurths_nature}
N.~Boers, B.~Goswami, A.~Rheinwalt, B.~Bookhagen, B.~Hoskins, J.~Kurths, Complex networks reveal global pattern of extreme-rainfall teleconnections, Nature 566 (2019) 373–377.

\bibitem{corral}
A.~Corral, Scaling in the timing of extreme event, Chaos, Solitons \& Fractals 74 (2015) 99–112.

\bibitem{Lacasa2008}
L.~Lacasa, B.~Luque, F.~Ballesteros, J.~Luque, J.~C. Nu{\~{n}}o, {From time series to complex networks: The visibility graph}, Proceedings of the National Academy of Sciences of the United States of America 105~(13) (2008) 4972--4975.
\newblock \href {https://doi.org/10.1073/pnas.0709247105} {\path{doi:10.1073/pnas.0709247105}}.

\bibitem{luque2009horizontal}
B.~Luque, L.~Lacasa, F.~Ballesteros, J.~Luque, Horizontal visibility graphs: Exact results for random time series, Physical Review E—Statistical, Nonlinear, and Soft Matter Physics 80~(4) (2009) 046103.

\bibitem{lacasa2009visibility}
L.~Lacasa, B.~Luque, J.~Luque, J.~C. Nuno, The visibility graph: A new method for estimating the hurst exponent of fractional brownian motion, EPL (Europhysics Letters) 86~(3) (2009) 30001.

\bibitem{lacasa2010description}
L.~Lacasa, R.~Toral, Description of stochastic and chaotic series using visibility graphs, Physical Review E—Statistical, Nonlinear, and Soft Matter Physics 82~(3) (2010) 036120.

\bibitem{lacasa2012time}
L.~Lacasa, A.~Nunez, {\'E}.~Rold{\'a}n, J.~M. Parrondo, B.~Luque, Time series irreversibility: a visibility graph approach, The European Physical Journal B 85~(6) (2012) 217.

\bibitem{lacasa2018visibility}
L.~Lacasa, W.~Just, Visibility graphs and symbolic dynamics, Physica D: Nonlinear Phenomena 374 (2018) 35--44.

\bibitem{lacasa2015time}
L.~Lacasa, R.~Flanagan, Time reversibility from visibility graphs of nonstationary processes, Physical Review E 92~(2) (2015) 022817.

\bibitem{iacovacci2016sequential}
J.~Iacovacci, L.~Lacasa, Sequential visibility-graph motifs, Physical Review E 93~(4) (2016) 042309.

\bibitem{luque2017canonical}
B.~Luque, L.~Lacasa, Canonical horizontal visibility graphs are uniquely determined by their degree sequence, The European Physical Journal Special Topics 226~(3) (2017) 383--389.

\bibitem{bohmer2024time}
T.~B{\"o}hmer, J.~P. Gabriel, L.~Costigliola, J.-N. Kociok, T.~Hecksher, J.~C. Dyre, T.~Blochowicz, Time reversibility during the ageing of materials, Nature Physics 20~(4) (2024) 637--645.

\bibitem{iacobello2021large}
G.~Iacobello, L.~Ridolfi, S.~Scarsoglio, Large-to-small scale frequency modulation analysis in wall-bounded turbulence via visibility networks, Journal of fluid mechanics 918 (2021) A13.

\bibitem{prl_cris}
A.~Aragoneses, L.~Carpi, N.~Tarasov, D.~V. Churkin, M.~C. Torrent, C.~Masoller, S.~K. Turitsyn, Unveiling temporal correlations characteristic to phase transition in the intensity of fibre laser radiation, Phys. Rev. Lett. 116 (2016) 033902.

\bibitem{sannino2017visibility}
S.~Sannino, S.~Stramaglia, L.~Lacasa, D.~Marinazzo, Visibility graphs for fmri data: Multiplex temporal graphs and their modulations across resting-state networks, Network Neuroscience 1~(3) (2017) 208--221.

\bibitem{zhuang2014time}
E.~Zhuang, M.~Small, G.~Feng, Time series analysis of the developed financial markets’ integration using visibility graphs, Physica A: Statistical Mechanics and its Applications 410 (2014) 483--495.

\bibitem{moraes2023b}
J.~T. Moraes, S.~C. Ferreira, Visibility graphs of critical and off-critical time series for absorbing state phase transitions, Physical Review E 108~(4) (2023) 044309.
\newblock \href {https://doi.org/10.1103/PhysRevE.108.044309} {\path{doi:10.1103/PhysRevE.108.044309}}.

\bibitem{Moraes2025}
J.~T. Moraes, S.~C. Ferreira, Strong localization blurs the criticality of time series for spreading phenomena on networks, Phys. Rev. E 111 (2025) 044302.
\newblock \href {https://doi.org/10.1103/PhysRevE.111.044302} {\path{doi:10.1103/PhysRevE.111.044302}}.

\bibitem{zhang2024spatio}
P.~Zhang, E.~Dai, C.~Wu, J.~Hu, Spatio-temporal patterns of hot extremes in china based on complex network analysis, Climate dynamics 62~(2) (2024) 841--860.

\bibitem{zhao2023visibility}
D.~Zhao, X.~Yang, W.~Song, W.~Zhang, D.~Huang, Visibility graph analysis of the sea surface temperature irreversibility during el ni{\~n}o events, Nonlinear Dynamics 111~(18) (2023) 17393--17409.

\bibitem{beltramone2025detecting}
G.~Beltramone, A.~C. Frery, M.~C. Scavuzzo, M.~Bonansea, A.~Ferral, Detecting seasonal snow transitions in sar time series with horizontal visibility graphs, Remote Sensing Applications: Society and Environment (2025) 101772.

\bibitem{zhang2022automated}
X.~Zhang, E.~C. Landsness, W.~Chen, H.~Miao, M.~Tang, L.~M. Brier, J.~P. Culver, J.-M. Lee, M.~A. Anastasio, Automated sleep state classification of wide-field calcium imaging data via multiplex visibility graphs and deep learning, Journal of neuroscience methods 366 (2022) 109421.

\bibitem{xuan2022avgnet}
Q.~Xuan, J.~Zhou, K.~Qiu, Z.~Chen, D.~Xu, S.~Zheng, X.~Yang, Avgnet: Adaptive visibility graph neural network and its application in modulation classification, IEEE Transactions on Network Science and Engineering 9~(3) (2022) 1516--1526.

\bibitem{aslan2024visgin}
H.~{\.I}. Aslan, C.~Choi, Visgin: Visibility graph neural network on one-dimensional data for biometric authentication, Expert Systems with Applications 237 (2024) 121323.

\bibitem{belhadi2025enhanced}
A.~Belhadi, P.~G. Lind, Y.~Djenouri, A.~Yazidi, Enhanced visibility graph for eeg classification, Frontiers in neuroscience 19 (2025) 1541062.

\bibitem{Boccaletti_2006}
S.~Boccaletti, V.~Latora, Y.~Moreno, M.~Chavez, D.-U. Hwang, Complex networks: Structure and dynamics, Phys. Reports 424 (2006) 175--308.

\bibitem{Boccaletti_2016}
M.~Zanin, D.~Papo, P.~Sousa, E.~Menasalvas, A.~Nicchi, E.~Kubik, S.~Boccaletti, Combining complex networks and data mining: Why and how, Phys. Reports 635 (2016) 1--44.

\bibitem{kantz2003nonlinear}
H.~Kantz, T.~Schreiber, Nonlinear time series analysis, Cambridge university press, 2003.

\bibitem{review_kurths}
Y.~Zou, R.~V. Donner, N.~Marwan, J.~F. Donges, J.~Kurths, Complex network approaches to nonlinear time series analysis, Phys. Reports 787 (2019) 1--97.

\bibitem{zhao2020extreme}
X.~Zhao, Y.~Li, J.~Wang, Extreme events analysis of non-stationary time series by using horizontal visibility graph, Fractals 28~(8) (2020) 2050150.
\newblock \href {https://doi.org/10.1142/S0218348X20500899} {\path{doi:10.1142/S0218348X20500899}}.

\bibitem{zhang2023extreme}
J.~Zhang, X.~Chen, H.~Wang, C.~Gu, H.~Yang, Visibility graph approach to extreme event series, Chinese Physics B 32~(10) (2023) 100505.
\newblock \href {https://doi.org/10.1088/1674-1056/acd62b} {\path{doi:10.1088/1674-1056/acd62b}}.

\bibitem{katsouda2025extreme}
M.~Katsouda, B.~Boutsinas, Detecting extreme values in time series based on bar visibility, International Journal of Data Science and Analytics (2025).
\newblock \href {https://doi.org/10.1007/s41060-025-00758-3} {\path{doi:10.1007/s41060-025-00758-3}}.

\bibitem{NOAA_Nino34_web}
{NOAA Physical Sciences Laboratory (PSL)}, {Climate Prediction Center (CPC) Niño 3.4 Index}, \url{https://psl.noaa.gov/data/timeseries/month/DS/Nino34_CPC/}, accessed 05-11-2026.

\bibitem{Rayner2003}
N.~A. Rayner, D.~E. Parker, E.~B. Horton, C.~K. Folland, L.~V. Alexander, D.~P. Rowell, E.~C. Kent, A.~Kaplan, Global analyses of sea surface temperature, sea ice, and night marine air temperature since the late nineteenth century, Journal of Geophysical Research: Atmospheres 108~(D14) (2003) 4407.
\newblock \href {https://doi.org/10.1029/2002JD002670} {\path{doi:10.1029/2002JD002670}}.

\bibitem{GISTEMPv4_2026}
{GISTEMP Team}, {GISS Surface Temperature Analysis (GISTEMP), Version 4}, \url{https://data.giss.nasa.gov/gistemp/}, accessed 05-12-2026 (2026).

\bibitem{NOAA_gmsst_web}
{NOAA Physical Sciences Laboratory (PSL)}, {Global Mean Land/Ocean Temperature Index}, \url{https://psl.noaa.gov/data/climateindices/list/}, accessed 05-11-2026.

\bibitem{Lenssen2024_noaa}
N.~Lenssen, G.~A. Schmidt, M.~Hendrickson, P.~Jacobs, M.~Menne, R.~Ruedy, A {GISTEMPv4} observational uncertainty ensemble, Journal of Geophysical Research: Atmospheres 129~(17) (2024) e2023JD040179.
\newblock \href {https://doi.org/10.1029/2023JD040179} {\path{doi:10.1029/2023JD040179}}.

\bibitem{shumway2011timeseries}
R.~H. Shumway, D.~S. Stoffer, Time Series Analysis and Its Applications: With R Examples, 3rd Edition, Springer, 2011.

\bibitem{estrada2012structure}
E.~Estrada, The structure of complex networks: theory and applications, American Chemical Society, 2012.

\bibitem{latora2017complex}
V.~Latora, V.~Nicosia, G.~Russo, Complex networks: principles, methods and applications, Cambridge University Press, 2017.

\end{thebibliography}

\end{document}